\documentclass[aps,prx,twocolumn,superscriptaddress]{revtex4-2}
\usepackage{amsmath}
\usepackage{amssymb}
\usepackage{graphicx}
\usepackage{color}

\newcommand{\av}[1]{\langle #1 \rangle}
\newcommand{\vp}{\varphi}

\begin{document}
\title{Finite-density-induced motility and turbulence of chimera solitons}

\author{L. A. Smirnov}
\affiliation{Department of Control Theory, Research and Education Mathematical Center ``Mathematics for Future Technologies'',
Nizhny Novgorod State University, Gagarin Av. 23, 603022, Nizhny Novgorod, Russia}
\author{M. I. Bolotov}
\affiliation{Department of Control Theory, Research and Education Mathematical Center 
``Mathematics for Future Technologies'',
Nizhny Novgorod State University, Gagarin Av. 23, 603022, Nizhny Novgorod, Russia}
\author{D. I. Bolotov}
\affiliation{Department of Control Theory, Research and Education Mathematical Center 
``Mathematics for Future Technologies'',
Nizhny Novgorod State University, Gagarin Av. 23, 603022, Nizhny Novgorod, Russia}
\author{G. V. Osipov}
\affiliation{Department of Control Theory, Research and Education Mathematical 
Center ``Mathematics for Future Technologies'',
Nizhny Novgorod State University, Gagarin Av. 23, 603022, Nizhny Novgorod, Russia}
\author{A. Pikovsky}
\affiliation{Institute of Physics and Astronomy, Potsdam University, 14476 Potsdam-Golm, Germany}
\affiliation{Department of Control Theory, Research and Education Mathematical 
Center ``Mathematics for Future Technologies'',
Nizhny Novgorod State University, Gagarin Av. 23, 603022, Nizhny Novgorod, Russia}

\begin{abstract}
We consider a one-dimensional oscillatory medium with a  coupling through
a diffusive linear field. In the limit of fast diffusion this setup reduces to the 
classical Kuramoto-Battogtokh model. We demonstrate that for a finite diffusion stable
chimera solitons, namely localized synchronous domain in an infinite asynchronous 
environment, are possible. The solitons are stable also for finite density of oscillators,
but in this case they sway with a nearly constant speed. This finite-density-induced motility
disappears in the continuum limit, as the velocity of the solitons is inverse 
proportional to the density. A long-wave instability of the homogeneous
asynchronous state causes soliton turbulence, which appears as a sequence of soliton mergings 
and creations. As the instability of the asynchronous state becomes stronger,
this turbulence develops into a spatio-temporal intermittency.
\end{abstract}
\maketitle
\section{Introduction}
Chimera patterns is a fascinating object in coupled oscillatory systems, attracting much of attention.
They were discovered and theoretically explained 
by Kuramoto and Battogtokh \cite{Kuramoto-Battogtokh-02} (KB) around 20 years ago. Later, 
Abrams and Strogatz \cite{Abrams-Strogatz-04} coined the term chimera to stress an unusual situation, where
in a fully spatially symmetric setup, oscillators form a synchronous (coherent) and asynchronous (disordered)
domains. Further theoretical and experimental studies of chimeras are 
summarized in reviews~\cite{Panaggio-Abrams-15,Omelchenko-18,scholl2016synchronization}. 

A striking property of chimeras is that
they are highly nontrivial at different levels of description, from microscopic through mesoscopic 
to macroscopic. At a \textit{microscopic} level one deals with a set of equations for coupled oscillators. In the simplest
KB setup it is a finite lattice (typically an ordered lattice, while recently
also lattices with a quenched or a time-dependent disorder have been 
studied~\cite{Smirnov_etal-21}) of phase oscillators with a non-local coupling with an exponential kernel.
Correspondingly, the observed regime is an attractor in a finite set of ordinary differential equations. A part of oscillators
form a synchronous cluster, while other units have frequencies different from the synchronous one. This order-disorder pattern can be classified as a weak chaos that is, for a large number of elements, close to a high-dimensional
quasiperiodic (with a large number of incommensurate frequencies) regime. 

One can drastically 
simplify the description at a \textit{mesoscopic} level in the continuous limit (which for a finite medium is also a 
thermodynamic limit). One introduces a coarse-grained order parameter, characterizing local distribution of oscillators with
one complex number. This order parameter obeys a nonlinear dissipative partial differential 
equation~\cite{Laing-09,Bordyugov-Pikovsky-Rosenblum-10,Smirnov-Osipov-Pikovsky-17}. This brings the problem
into realm of pattern formations in nonlinear nonequilibrium media~\cite{Pismen-06,Cross-Greenside-12}. In this context
the \textit{macroscopic} KB chimera is a stationary (in a proper rotating system) 
spatially periodic pattern~\cite{Omelchenko_etal-08,Laing-09,Omelchenko_etal-10,Laing-11,Omelchenko-13,Maistrenko_etal-14,Xie_etal-14,Xie_etal-15,Omelchenko-Knobloch-19}. This pattern may become unstable, what results in a breathing 
chimera~\cite{Kemeth_etal-16,Suda-Okuda-18,Bolotov_etal-17a,Bolotov-Smirnov-Osipov-Pikovsky-18,Omelchenko-20} 
or in turbulent regimes~\cite{Bordyugov-Pikovsky-Rosenblum-10}. In this description the difference between synchronous and asynchronous domains
is not as astonishing as at the microscopic level; here local synchrony and asynchrony (or, in other words,
local order and local disorder) are determined just by the
values of the local complex order parameter (which has absolute value one for synchrony, and is less than one
for asynchrony and partial synchrony).

In this paper we study chimera solitons. At the mesoscopic and the macroscopic levels, 
existence of solitary solutions of dissipative
partial differential equations is not surprizing, but previous attempts to find such solutions resulted in unstable
solitons~\cite{Smirnov-Osipov-Pikovsky-17}. The crucial step here is to restore a physical description
of the coupling which was at the origin of the KB model. Indeed, an exponential interaction kernel of coupling between
the oscillators is motivated by a physical situations where these oscillators interact via a diffusive field. The 
instantaneous exponentially decaying coupling appears in the limit of very fast diffusion. 
Below we consider a more realistic situation of finite diffusion and find that this facilitates stability of the 
solitary chimeras, which can be described as standing local synchronous domains on an 
infinite disordered background. It belongs to a class of dissipative 
solitons~\cite{Ackemann-09,Purwins-10,Kerner-Osipov-13} (thus, in constradistiction to solitons in conservative nonlinear wave equations, its parameters do not depend on initial conditions). 
Furthermore, at the same macroscopic level we study a situation where the
homogeneous disordered state becomes unstable. Here the solitons appear spontaneously, but they do not form
a regular stable lattice - instead we observe an interesting regime of soliton turbulence, where solitons merge 
and emerge from the background in an irregular manner. Chimera solitons have a bounded amplitude 
(because the order parameter cannot exceed one), so if the stability of a synchronous state is 
enhanced, at merging events wide synchronous domains appear. We demonstrate that this leads to 
a spatio-temporal intermittency~\cite{Chate-Manneville-87}, 
where the synchronous state  is the absorbing one.

The found solitons can be straightforwardly modelled on a microscopic level, i.e. in a lattice of oscillators
coupled via a diffusive field. These simulations reveal a rather unexpected effect: for a  finite density of oscillators, solitons
start to move. They move nearly regularly, but from time to time switch the direction, so that 
the whole dynamics looks like swayings. We call this phenomenon \textit{finite-density-induced motility} 
because it disappears in the continuum limit.
We argue that this motility is due to finite-size fluctuations, and in this sense belongs to the class of effects where
these fluctuations play a constructive role (cf. finite-size-induced 
transitions~\cite{Pikovsky-Rateitschak-Kurths-94,Komarov-Pikovsky-15b}
and system size resonance \cite{Pikovsky-Zaikin-Casa-02}). 

The paper is organized as follows. In Sec.~\ref{sec:bm} we introduce the model of a one-dimensional
medium of oscillators coupled through a diffusive field, including the mesoscopic description
in terms of the local order parameter. In Sec.~\ref{sec:hss} we briefly discuss spatially homogeneous
states and their stability.  In Sec.~\ref{sec:solchim} we formulate the problem of finding solitary states.
We present an approximate analytical approach and compare its results with numerical ones.
In Sec.~\ref{sec:fsim} we explore chimera solitons in systems with a finite density of oscillators
and demonstrate the effect of finite-size-induced motility. We show, in particular, that the velocity decreases inversely proportional to density. In Sec.~\ref{sec:st} we return to the mesoscopic description with
partial differential equations, and explore soliton turbulence which appears when the 
disordered homogeneous state becomes unstable. We summarize the results in Sec.~\ref{sec:concl}.

\section{Basic model}
\label{sec:bm}
\subsection{Phase dynamics}
We consider a one-dimensional medium of oscillators which interact  not directly, but through a diffusive
``chemical agent field''. This class of models has been introduced by Y.~Kuramoto and co-workers
\cite{Kuramoto_etal-00,Kuramoto-Battogtokh-02,Tanaka-Kuramoto-03,Shima-Kuramoto-04}. 
The oscillators are described by their phases $\vp(x,t)$, and the medium by a complex field $P(x,t)$,
which is dissipative and diffusive, and is driven by the oscillators.
The equations have the form
\begin{equation}
\begin{aligned}
\partial_{t}\varphi&=\zeta+\epsilon\text{Im}(P e^{-i\alpha}e^{-i\varphi})\;,\\
\partial_{t}P&=A\partial_{xx}P-B P+C e^{i\varphi}\;.
\end{aligned}
\label{eq:or}
\end{equation}
Here parameter $\zeta$ is the frequency of oscillators; parameters 
$\epsilon,\alpha, C$ describe coupling between the oscillators and the chemical agent,
and parameters $A,B$ describe diffusion and dissipation of the latter. All parameters are real; the phase shift in the coupling is described by angle $\alpha$. It is convenient to reduce the number 
of variables by rescaling time, space, and the field amplitude $P$ according to
\[
t=a t',\qquad x=b x',\qquad P=cH
\]
with
\[
c={C}{B^{-1}},
\quad b^2={A}{B^{-1}},\quad a={B}{(\epsilon C)^{-1}}.
\]
Then the equations take the form (we omit primes at new time and space variables)
\begin{subequations}\label{eq:or1}
\begin{align}
\partial_{t}\varphi&=\omega+\text{Im}(H e^{-i\alpha}e^{-i\varphi})\;,\label{eq:or1a}\\
\tau\partial_{t}H&=\partial_{xx}H- H+ e^{i\varphi}\;.\label{eq:or1b}
\end{align}
\end{subequations}
These equations contains three essential parameters: dimensionless
frequency of oscillations $\omega=\zeta B (\epsilon C)^{-1}$, dimensionless 
relaxation time of the agent field $\tau=\epsilon C B^{-2}$, and the coupling phase
shift $\alpha$. Additionally, the length of the system is a parameter, but in this
paper we consider patterns in an infinite domain.

Eq.~\eqref{eq:or1b} has a remarkable limit of fast dynamics of the agent field $H$,
where one sets $\tau\to0$. In this limit the agent can be represented via the forcing
field as an integral 
\[
H(x,t)=\int e^{-|x'-x|} e^{i\vp(x',t)}\,dx'\;,
\]
substitution of which to the equation for the phases \eqref{eq:or1a} leads to an effective non-local coupling
with an exponential kernel. Note that transforming to the rotating reference frame
one can get rid of parameter $\omega$ as well. The resulting model has
been shown by Kuramoto and Battogtokh~\cite{Kuramoto-Battogtokh-02}
to possess chimera patterns in a periodic domain (for a similar result in two dimensions 
see \cite{Shima-Kuramoto-04}).

In this paper we do not assume the relaxation parameter $\tau$ to be small. We will demonstrate
that this allows for existence of stable chimera solitons in an infinite domain.

\subsection{Ott-Antonsen reduction}
While Eqs.~\eqref{eq:or1} can be straightforwardly discretized 
for numerical simulations (we perform this in Sec.~\ref{sec:fsim}),
their analytical treatment is hardly possible because the field $\vp$ is not smooth
in space -- neighboring phases are (almost) independent in a desynchronized domain.
For a tractable system of partial differential equations (PDEs) one performs
a local Ott-Antonsen (OA) reduction (see \cite{Laing-09,Bordyugov-Pikovsky-Rosenblum-10} for its application to such a setup).
It is based on a local averaging over the distribution of the phases; the latter
is assumed to be a wrapped Cauchy distribution which is characterized by a single complex number 
$Z(x,t)$ -- the local Kuramoto order parameter $Z(x,t)=\av{e^{i\vp}}_{\text{loc}}$, where
the averaging is performed over a small neighborhood of a site $x$. Accordingly, the OA 
reduction requires a thermodynamic limit, where the density of the oscillators
tends to infinity.
The Eqs.~\eqref{eq:or1} after the OA reduction take the form
\begin{subequations}
\begin{align}
\partial_t Z&=i\omega Z+\frac{1}{2}\left(e^{-i\alpha}H-e^{i\alpha}H^* Z^2\right)\;,
\label{eq:oaz}\\
\tau \partial_t H&=\partial_{xx}H-H+Z\;.
\label{eq:oah}
\end{align}
\label{eq:oa}
\end{subequations}
Now, one can consider continuous profiles of order parameter $Z$ (the profiles of $H$ are anyhow relatively smooth due to diffusion). The order parameter $Z$ reaches it maximal possible value
$|Z|=1$ in synchronous domains, and $|Z|<1$ in partially synchronous regions (the full asynchrony with
a uniform distribution of phases corresponds to $|Z|=0$).

\section{Homogeneous states and their stability}
\label{sec:hss}
\subsection{Homogeneous states}
\begin{figure}[!t]
\centering
\includegraphics[width=1.0\columnwidth]{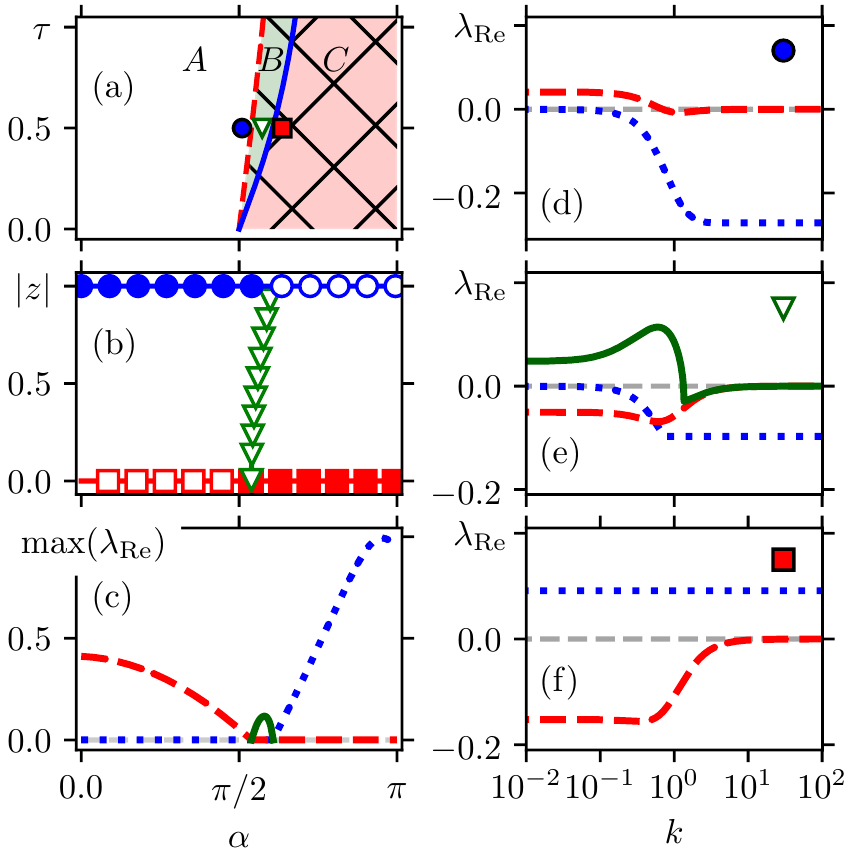}
\caption{Existence and stability of spatially homogeneous states.
Panel (a): Domains of existence and stability of homogeneous states on the plane $(\alpha,\tau)$
for $\omega=0.25$. The fully asynchronous state (FAS) is stable in domains B and C and unstable in domain A;
the completely synchronous state (CSS) is stable in domains A and B and unstable in domain C; the partially synchronous 
state (PSS) exists in domain B and is unstable. Panel (b): The order parameter $|z|$ of the 
homogeneous regimes in dependence on $\alpha$ for $\omega=0.25$, $\tau=0.5$. 
Red squares -- FAS, blue circles -- CSS, green triangles -- PSS. Filled (empty) markers denote stable (unstable) 
regimes. Panel (c): Maximum (over all wavenumbers) real part of the mostly unstable
eigenvalue $\max \left( \lambda_{\mathrm{Re}} \right)$ of homogeneous states for $\omega=0.25$, $\tau=0.5$. 
Red dashed line -- FAS, blue dotted line -- CSS, green solid line -- PSS. Panels (d, e, f): Real parts of the 
mostly unstable eigenvalue $\max \left( \lambda_{\mathrm{Re}}(\kappa) \right)$ for 
$\omega=0.25$, $\tau=0.5$ and $\alpha = 1.6$ (d), $\alpha = 1.8$ (e), $\alpha = 2.0$ (f).}
\label{fig:om}
\end{figure}

For $\tau=0$, i.e. in the standard KB setup, Eqs.~\eqref{eq:oa} have two 
spatially homogeneous solutions: one fully asynchronous with $Z=0$, and one
fully synchronous with $|Z|=1$. These states exchange stability at $\alpha=\pi/2$,
i.e. exactly at the value of the phase shift that corresponds to a neutral coupling.
With $\tau,\omega\neq 0$, another homogeneous state with partial synchrony $0<|Z|<1$ becomes possible.
We will not discuss here the full bifurcation diagram for homogeneous
solutions, but mention only that there is a range of relatively small values of $\tau,\omega$,
where the homogeneous states have the following properties (see Fig.~\ref{fig:om}). 
There is a range of values
of the phase shift parameter $\alpha_l<\alpha<\alpha_r$, where the partially
synchronous state exists and is unstable. It coexists with the stable fully asynchronous 
and stable fully synchronous states. The asynchronous state becomes unstable for $\alpha<\alpha_l$,
while the synchronous state becomes unstable for $\alpha>\alpha_r$.

Analytically, these properties are derived as follows. We seek for
uniformly rotating solutions  and substitute 
\begin{equation}
H=h_0 e^{i(\omega+\Omega) t},\qquad Z = z_0 e^{i(\omega+\Omega) t}
\end{equation}
in Eq.~\eqref{eq:oa}. 
Then $h=(1+i\tau(\omega+\Omega))^{-1}z_0$ and the equation for $z_0$ reads
\begin{equation}
0=z_0\left[-2i\Omega+e^{-i\alpha}\frac{1}{1+i\tau\Omega}-
e^{i\alpha}\frac{|z_0|^2}{1-i\tau\Omega}\right].
\label{eq:sus1}
\end{equation}
One can see that for all parameter values $\omega,\tau,\alpha$ there are at least
two solutions: one fully asynchronous with $|z|=z_{fa}=0$, and one fully synchronous
with $|z|=|z_{fs}|=1$. Additionally, in some ranges of parameters, a partially synchronous
solution with $0<|z|<1$ exists. This nontrivial solution has frequency 
$\Omega_{ps}=\tau^{-1}\cos\alpha/\sin\alpha-\omega$, its amplitude follows from the relation
\[
|z_{ps}|^2=\frac{2(\tau\omega\sin\alpha-\cos\alpha)}{\tau\sin^2\alpha}-1\;.
\]
Additionally, the condition $0 < |z_{ps}| < 1$ should be satisfied. 
Note that for $\tau=0$ a partially synchronous state can exist
for $\cos\alpha=0$ only, what corresponds to a conservative coupling. In Fig.~\ref{fig:om}
we show regions of existence of the partially synchronous state for several values of $\omega$ and for
$\tau<1$ (in a larger range of parameters these domains become more complex, but it is not our aim here to provide a full analysis). Panel (a) of  Fig.~\ref{fig:om} shows domains of existence of different states for $\omega=0.25$
on the plane $(\alpha,\tau)$. Panel (b) shows the values of $|z|$ for these states at $\tau=0.5$.

\subsection{Stability of homogeneous states}
A detailed stability analysis of homogeneous states is presented in Ref.~\cite{Bolotov_etal-21}, here
we discuss briefly only the properties relevant for the soliton dynamics below. It is instructive to start with
the case $\tau=0$. Then the stability properties are determined solely by parameter $\alpha$, which governs
the nature of coupling: there is a change from
attractive to repulsive coupling at $\alpha=\pi/2$. Correspondingly, the fully unstable and the fully stable
states exchange their stability at $\alpha=\pi/2$. For small $\tau$ and $\omega$ this picture slightly changes:
a domain $\alpha_l<\alpha<\alpha_r$ appears, where both these states are stable; this is exactly
the domain where also a partially synchronous state exists, which is unstable. This is illustrated in Fig.~\ref{fig:om}(c),
where the maximal (over the full range of wavenumbers $k$) values of the growth rate of
small perturbations on top of homogeneous states are shown. The nature of stability and instability is
clear from panels (d-f), where we depict the real part of the eigenvalues in dependence on the wavenumber $k$,
at three parameter values denoted with the markers. In the situation shown in panel (e),
 both states $|z_{fs}|=1$ and $|z_{fa}|=0$ are stable, but the intermediate state is unstable.  
For the soliton solution below the stability properties of the fully asynchronous state are mostly
important. There are two possible types of instability: one is a long-wave instability (unstable are modes
with wave numbers $0\leq k \leq k_{lw}$), another is a short-wave instability (modes
with $k\geq k_{sw}$ are unstable). For $\omega<1/2$ only the long-wave instability is relevant.
This case is illustrated in Fig.~\ref{fig:om}(d).

\section{Solitary chimera}
\label{sec:solchim}
\subsection{Equations for a solitary profile}

We look for solitary solutions of Eqs.~\eqref{eq:oa} in the form
\begin{equation}
	Z(x,t)=z(x) e^{i(\omega-\nu) t},\quad H(x,t)=h(x) e^{i(\omega-\nu) t}.
	\label{eq:ZH}
\end{equation}
The value of $\nu$ is the ``eigenvalue'' of the found solution,
for $\alpha$ close to $\pi/2$ it is positive. Then Eq.~\eqref{eq:oaz} reduces to an algebraic equation
$-2i\nu z=e^{-i\alpha}h-e^{i\alpha}h^*z^2$. Only the solution having
absolute value less than one should be used:
\begin{equation}
z(x)=\frac{\nu-\sqrt{\nu^2-|h(x)|^2}}{e^{i\beta}h^*(x)}.
\label{eq:zh}
\end{equation}
Here we introduce $\beta=\alpha-\pi/2$, it will serve as a small parameter
in the analytic expressions for a soliton. Substituting \eqref{eq:zh} in \eqref{eq:oah},
we obtain a differential equation for the complex field $h(x)$:
\begin{equation}
h''-\big(1+i\tau(\omega-\nu)\big)h=\frac{\sqrt{\nu^2-|h|^2}-\nu}{e^{i\beta}h^*}.
\label{eq:solh}
\end{equation}
We seek for localized solutions with $|h(x)|\to 0$ for $x\to\pm\infty$, i.e.,
for homoclinic solutions of \eqref{eq:solh}.

Both for a numerical analysis and for a theoretical treatment it is convenient to rewrite
the fourth-order ODE Eq.~\eqref{eq:solh} as a third-order system \cite{Smirnov-Osipov-Pikovsky-17},
this is always possible due to invariance of these equations to phase shifts. Representing
$h(x)=r(x)\exp[i\theta(x)]$ and introducing a variable $q=r^2\theta'$, we get a
system of real equations which have different form at small and large amplitudes $r$.
In domain $|r|\leq |\nu|$ (this corresponds to a locally partially synchronous state of
oscillators) the system reads
\begin{subequations}
\begin{align}
r''&=r+\frac{q^2}{r^3}+\frac{\sqrt{\nu^2-r^2}-\nu}{r}\cos\beta,
\label{eq:smallrr}\\
q'&=\tau(\omega-\nu)r^2-(\sqrt{\nu^2-r^2}-\nu)\sin\beta.
\label{eq:smallrq}
\end{align}
\label{eq:smallr}
\end{subequations}
In domain $|r|> |\nu|$ (this corresponds to a synchronous state of
oscillators) the system reads
\begin{subequations}
\begin{align}
r''&=r+\frac{q^2}{r^3}-\frac{\nu}{r}\cos\beta + \frac{\sqrt{r^2-\nu^2}}{r}\sin\beta,
\label{eq:largerr}\\
q'&=\tau(\omega-\nu)r^2+\nu\sin\beta+\sqrt{r^2-\nu^2}\cos\beta.
\label{eq:largerq}
\end{align}
\label{eq:larger}
\end{subequations}
A homoclinic solution (with $r,q$ vanishing at $\pm\infty$), which completely 
lies in the domain $|r|< |\nu|$ 
is a partially synchronous soliton (it has no synchronous part);
 a homoclinic solution with a patch in the domain   $|r|> |\nu|$
is a chimera soliton, it has a central synchronous part and disordered tails.

A numerical procedure for finding a soliton is straightforward: one starts two solutions in a vicinity
of $r=q=0$ at large positive and large negative values of $x$, and by virtue of shooting matches these
solutions at $x=0$ using $\nu$ as the adjustment parameter. In this way, for different
parameters $\tau,\omega,\beta$ one finds families of solitons. However, there is a possibility
to construct an analytic approximation to a soliton solution, which we present in Sec.~\ref{sec:ansol}.

\subsection{Analytic approximation of a soliton and comparison to numeric}
\label{sec:ansol}
As has been first explored in \cite{Smirnov-Osipov-Pikovsky-17},
for $\beta=\tau=0$ one can describe the partially synchronous soliton analytically.
Indeed, in this case $q=0$ is a solution of \eqref{eq:smallrq}, and \eqref{eq:smallrr}
is a second-order equation, solution of which can be represented as an integral. We will show below
that for small $\beta,\tau$ the branches of solitons can be found in a
perturbative manner.

It is convenient to set first $q=0$ in both Eqs.~\eqref{eq:smallrr} and 
\eqref{eq:largerr} (although in the latter case it is not an exact solution 
of \eqref{eq:larger} even for $\beta=0$).
This allows for introducing a potential $U(r)$ and for formulation of 
the Eqs.~\eqref{eq:smallrr} and 
\eqref{eq:largerr}
as the dynamics in this potential (here and below, for brevity of notations, we use that $\nu>0$):
\begin{widetext}
\begin{equation}
r'' =-\frac{dU(r)}{dr},\qquad 	U(r) =
	\begin{cases}
		\displaystyle -{r^2}/{2} - \sqrt{\nu^2 - r^2} + \nu \ln\big({\sqrt{\nu^2 - r^2} +\nu}\big)-\nu \big(\ln({2\nu}) - 1 \big), &|r| < \nu, \\
		\displaystyle -{r^2}/{2} + \nu \ln\big({|r|}\big)-\nu \big(\ln\big({2\nu}\big) - 1 \big), &|r| \geq \nu.
	\end{cases}
	\label{eq:pot}
\end{equation}
\end{widetext}
Here a constant term is added to $U(r)$ to ensure $U(0)=0$. 

\begin{figure}[!t]
	\centering
	\includegraphics[width=1.0\columnwidth]{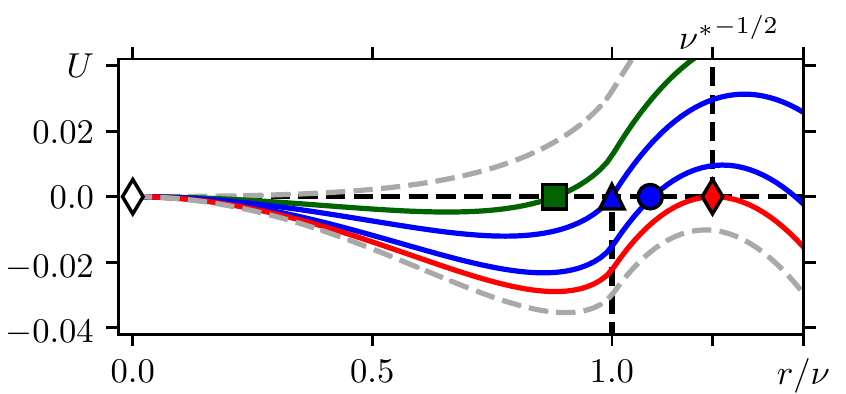}
	\caption{Potential function $U(r)$ \eqref{eq:pot}
	for different values of the frequency $\nu$. Green solid line with a square marker corresponds to 
	a soliton regime without a fully synchronous domain for $ \nu^{***}<\nu<\nu^{**}$. Blue solid line with 
	a triangular marker corresponds to a one-point soliton chimera for $\nu=\nu^{**}$. Blue solid line with a circle 
	marker corresponds to a soliton chimera for $\nu^{**}<\nu<\nu^{*}$. Red solid line with a diamond marker 
	corresponds to a soliton with 
maximal amplitude for $\nu=\nu^{*}$. Gray dashed lines 
correspond to non-soliton solutions. }
	\label{fig:pot}
\end{figure}

Let us now discuss possible soliton solutions of \eqref{eq:pot} (these are generally not the soliton
solutions of the original equations because we assume $q=0$ here).
Potential \eqref{eq:pot} (see Fig.~\ref{fig:pot}) has a maximum  at $r=0$ for $\nu >\nu^{***}=0.5$,
this defines one border for the soliton existence. The potential has also a maximum at 
finite amplitude $r^*=(\nu^{*})^{1/2}$. The condition  $U(r^*)=0$ defines another border for the soliton 
existence: $\nu^{*}=e/4\approx 0.67957$. The change of the type of  soliton (i.e. whether 
it is completely in the domain $r<\nu$
or has a part in the domain $r\geq \nu$ happens if $U(r^{**})=0$ and $r^{**}=\nu^{**}$.
This condition yields $\nu^{**}=2(1-\ln 2)\approx 0.6135$. 
Only solitons for $ \nu^{***}<\nu<\nu^{**}$, which are fully partially synchronous, are also
the solitons of the full equations \eqref{eq:smallr} for $\tau=\beta=0$ (because in this case also $q=0$).
In contradistinction, solitons in \eqref{eq:pot} for $\nu^{**}<\nu<\nu^{*}$ are not exact solutions
of \eqref{eq:smallr},\eqref{eq:larger}, but nevertheless they can serve as initial approximations 
in the perturbation method below.

Next we perform a perturbation analysis, assuming that the variable $q$ (which above has been assumed to vanish) 
remains small.
This is ensured if parameters $\tau(\omega-\nu)$ and $\beta$ are small. Additionally, 
the chimera soliton should have a small synchronous domain, i.e. $\sqrt{r^2-\nu^2}$ remains small
(practically this means that we have to consider $\nu$ close to $\nu^{**}$, this is required to keep the last term
in Eq.~\eqref{eq:largerq} small).
Under these conditions, the equation for the variable $r$ remains the same Eq.~\eqref{eq:pot} (additional terms
in Eqs.~\eqref{eq:smallrr} and \eqref{eq:largerr} are of second order in small parameters). 
Therefore, we have only to ensure
that the condition $q\to 0$ at $x\to\pm\infty$ is fulfilled. The equation for $q$ (here we combine Eqs.~\eqref{eq:smallrq} and \eqref{eq:largerq}) reads
\begin{equation}
	q' = g_1(r)-\beta g_2(r),
\label{eq:qeq}
\end{equation}
where  we introduced two functions (containing the Heaviside step function $\Theta$)
\begin{align}
		g_1(r)& = \tau (\omega -\nu) r^2 + \Theta(r - \nu) \sqrt{r^2 - \nu^2},\\
		g_2(r) &= -\nu + (1 - \Theta(r - \nu)) \sqrt{\nu^2 - r^2}\;.    
\end{align}
Because $q(x)$ is an odd function of $x$, it should vanish at $x=0$, and 
the condition that $q\to 0$ at $x\to\pm\infty$ 
can be achieved by integrating Eq.~\eqref{eq:qeq} in the domain $(-\infty,0)$:
\[
\int_{-\infty}^0 (g_1(r(x))-\beta g_2(r(x)))dx=0.
\]
 It is convenient to rewrite the integral in terms of variable $r$, using $r'=\sqrt{-2 U(r)}$. This gives
 the family of soliton solutions in form of dependence of parameter $\beta$ on the parameters
 of the problem $\tau,\omega$ and on the frequency of the solution $\nu$:
 \begin{equation}
 \beta =\frac{\displaystyle\int\limits_0^{r_{max}} \displaystyle\frac{g_1(r)}{\sqrt{-2U(r)}} dr}{\displaystyle\int\limits_0^{r_{max}} \displaystyle\frac{ g_2(r) }{ \sqrt{-2 U(r)} } dr}.
\label{eq:betapt}
\end{equation}
Here $r_{max}$ is defined from the condition $U(r_{max})=0$,
according to Eq.~\eqref{eq:pot} this quantity is a function of $\nu$.
In the case of a soliton which is partially synchronous, the integration is only over domain $r_{max}<|\nu|$
and the expression \eqref{eq:betapt}
simplifies to 
\begin{equation}
		\beta = \tau(\omega -\nu)\frac{\displaystyle\int\limits_0^{r_{max}} \displaystyle\frac{ r^2 }{ \sqrt{-2U(r)} } dr}{\displaystyle \int\limits_0^{r_{max}} 
		\displaystyle\frac{ -\nu + \sqrt{\nu^2 - r^2} }{ \sqrt{-2U(r)} } dr}.
	\label{eq:betaps}   
\end{equation}
In the case of a chimera soliton, one has to perform integration in \eqref{eq:betapt} in both domains
$r<\nu$ and $r>\nu$. 

Expressions \eqref{eq:betapt}, \eqref{eq:betaps} are the main result of the approximative analytical approach. On a qualitative level, they imply existence of two branches of solitons, one partial synchronous, and another chimera soliton.
For both these branches \eqref{eq:betapt}, \eqref{eq:betaps} yield dependencies of the ``eigenfrequency'' $\nu$
on the parameters $\beta, \tau, \omega$. 

\begin{figure*}[!t]
	\centering
	\includegraphics[width=2.0\columnwidth]{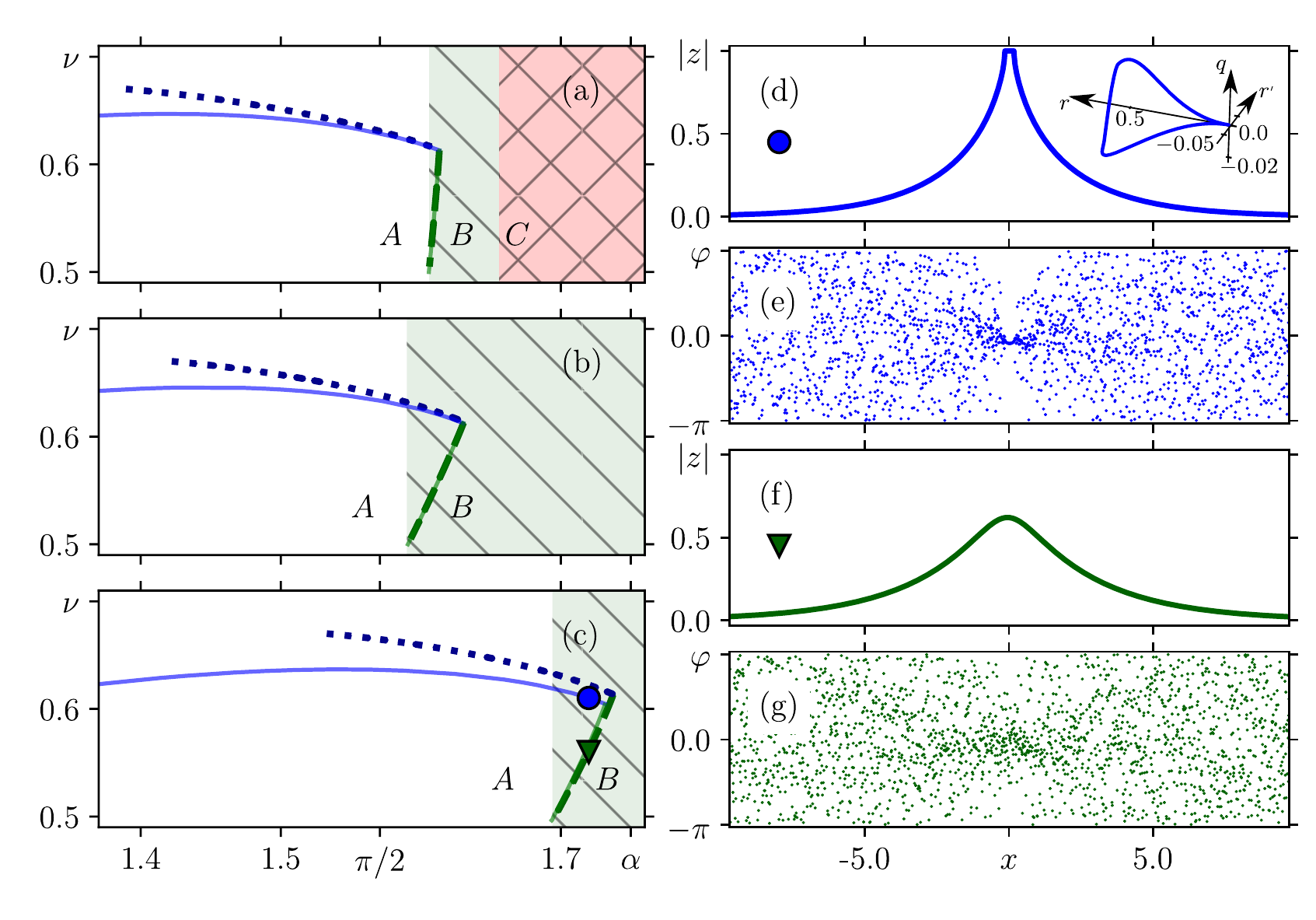}
	\caption{Panels (a, b, c): Dependencies of the parameter $\nu$ on $\alpha$ of the soliton solutions~\eqref{eq:ZH} for $\omega=0.15$, $\tau=0.1$ (a); $\omega=0.45$, $\tau=0.4$ (b); and $\omega=0.25$, $\tau=0.5$ (c). Dashed and dotted lines -- analytical results from Eqs.~\eqref{eq:betapt}, \eqref{eq:betaps}. Solid lines -- numerical results based on finding homoclinic trajectories of the system~\eqref{eq:smallr}, \eqref{eq:larger}. Blue lines -- soliton chimeras, green lines -- soliton solutions without a fully coherent patch.  
	FAS is stable in domains B and C and unstable in domain A; CSS is stable in domains A and B and unstable in domain C; PSS exists in domain B and is unstable. 
	Panels (d, e): Soliton chimera profiles for $\omega=0.25$, $\tau=0.5$, $\alpha=1.72$. Panels (f, g): Soliton solutions without a fully coherent domain for $\omega=0.25$, $\tau=0.5$, $\alpha=1.72$. Panels (d, f) show modulus of local order parameter $|z(x)|$, panels (e, g) show phases $\varphi(x)$, respectively. Inset on panel (d) depicts a homoclinic trajectory corresponding to 
	the  chimera soliton in the phase space of the system~\eqref{eq:smallr}, \eqref{eq:larger}.}
	\label{fig:ansol}
\end{figure*}

In Fig.~\ref{fig:ansol} we compare the numerically found chimera solitons with the analytic approximations \eqref{eq:betapt}, \eqref{eq:betaps}.
Because the perturbation term is $\sim\tau(\omega-\nu)$, and $\nu$ is positive, we observe a good correspondence of the numerical solutions and their analytical approximations both for small values of $\tau$ (panel (a))
and for moderate $\tau$ and large $\omega$ (panel (b)). The correspondence is not so good (although qualitatively
correct) for moderate $\tau$ and relatively small $\omega$ (panel (c)). Panels (d) and (f) show profiles
of the order parameter for a chimera soliton and for a partially synchronous soliton, correspondingly.
Panels (e,g) depict the corresponding snapshots of the phases for a finite-density representation of the solitons (see
Sec.~\ref{sec:fsim}).

\begin{figure}[!t]
	\centering
	\includegraphics[width=1.0\columnwidth]{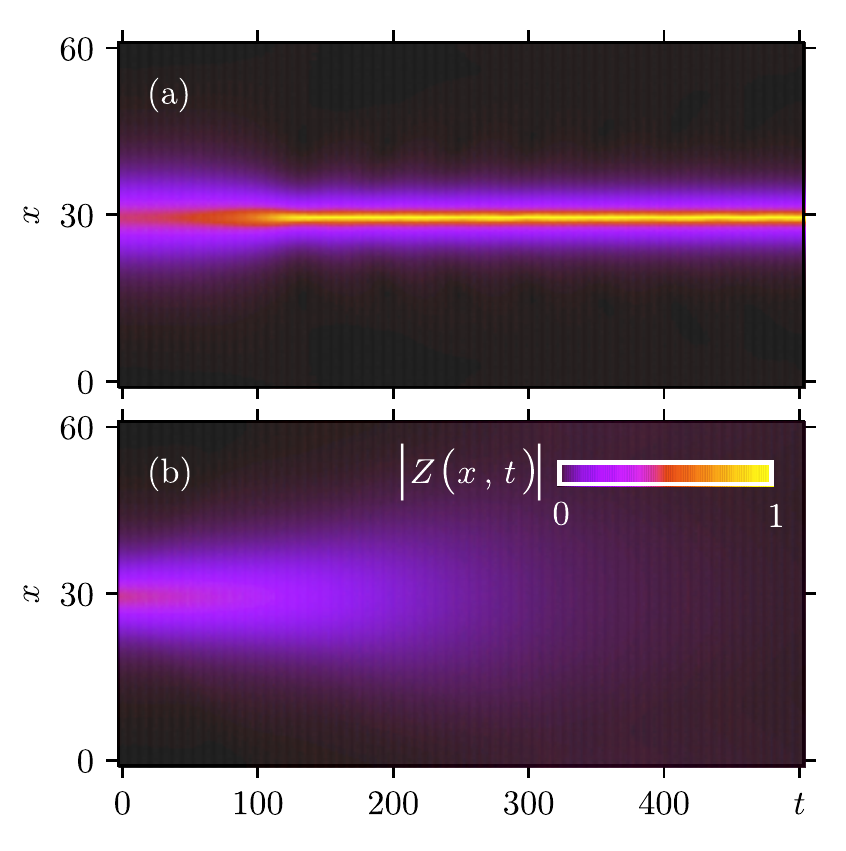}
	\caption{Evolution of numerically found solitary waves in simulations
	of the full equations \eqref{eq:oa} in a domain of length $L=60$. The color code
	shows the field $|z(x,t)|$. Parameters: $\tau=0.5$,
	$\omega=0.25$, $\alpha=1.71$. Panel (a):  partially synchronous soliton as an initial condition;
	because of instability a stable chimera soliton appears.
	Panel (b): the initial condition of panel (b) multiplied by factor $0.95$, here the instability results in a homogeneous asynchronous state. }
	\label{fig:solf}
\end{figure}

By virtue of direct numerical simulations (in a large $L\gg 1$ spatial domain with periodic boundary conditions)
we have found that chimera solitons (Fig.~\ref{fig:ansol}(d,e)) 
are stable 
in the domain B of Fig.~\ref{fig:ansol}, i.e. in the domain where the asynchronous homogeneous state
(on top of which the soliton exists) is stable. The partially synchronous solitons (Fig.~\ref{fig:ansol}(f,g))
are unstable. 
We illustrate this in Fig.~\ref{fig:solf}.
Here in panel (a) the initial condition is taken as a chimera soliton, and no evolution
is seen. In panel (b) we take a partially unstable soliton as an initial condition, and it evolves
toward the chimera soliton existing at the same values of parameters. In panel (c)
we take a partially unstable soliton multiplied by factor $0.95$ as an initial condition, it evolves
toward the stable homogeneous asynchronous state.

\section{Finite-density-induced motility}\label{sec:fsim}
In this section we report on numerical exploration of the found soliton
in a discrete lattice of oscillators. 

\subsection{Discrete equations}
We consider first a situation where the oscillators are placed at discrete positions $x_n$,
with spacing $\Delta x=x_{n+1}-x_n$, but the coupling field $H$ is continuous in space.
Then the system \eqref{eq:or1} can be rewritten as
\begin{subequations}\label{eq:or2}
\begin{align}
\dot{\varphi}_n&=\omega+\text{Im}(H(x_n,t) e^{-i\alpha}e^{-i\varphi_n})\;,\label{eq:or2a}\\
\tau \partial_t H&=\partial_{xx}H- H+ \Delta x \sum_n \delta(x-x_n)e^{i\varphi_n}\;.\label{eq:or2b}
\end{align}
\end{subequations}
For the next step, namely for discretization of the field $H$, one has to compare
a characteristic spatial scale of diffusion (which is in our case one) with the spacing
$\Delta x$. If $\Delta x \ll 1$, then the variations of $H$ on scale $\Delta x$ are small.
Thus, this field can be characterized by its coarse-grained values $H_n$ in intervals 
$(x_n-\Delta x/2,x_n+\Delta x/2)$:
\[
H_n(t)=\rho \int_{x_n-\Delta x/2}^{x_n+\Delta x/2} H(x',t) d x'\;,
\]
where we introduced the density $\rho=1/\Delta x$. Applying the integration over the 
interval $(x_n-\Delta x/2,x_n+\Delta x/2)$ to Eq.~\eqref{eq:or2b} 
and approximating the second derivative
via the values $H_{n-1},H_n,H_{n+1}$, we obtain a set of discrete equations 
suitable for numerical simulations:
\begin{subequations}\label{eq:or3}
\begin{align}
\dot{\varphi}_n&=\omega+\text{Im}(H_n e^{-i\alpha}e^{-i\varphi_n})\;,\label{eq:or3a}\\
\tau \dot{H}_n&=\rho^2(H_{n+1}-2 H_n+H_{n-1})- H_n+ e^{i\varphi_n}\;.\label{eq:or3b}
\end{align}
\end{subequations}
In all simulations below in this section, a finite spatial domain of length 
$L\gg 1$ with periodic boundary conditions
is used. The total number of oscillators is $\rho L$. 
Furthermore, we fix parameters at values $\alpha=1.71$, $\tau=0.5$, $\omega=0.25$. 

\subsection{Soliton motility at finite densities}
In discrete simulations of a soliton in the framework of system \eqref{eq:or3} we expect
to observe finite-size effects. The parameter governing these effects 
is not the number of oscillators (because formally the
solution is in an infinite domain), but density $\rho$. Because the characteristic
length of solitons reported in Sec.~\ref{sec:solchim} is one, parameter $\rho$ gives approximately
the number of fully synchronized oscillators. For large values of $\rho$, we expect the solution
of the discrete version~\eqref{eq:or3} to ``converge'' to that of the continuous equations~\eqref{eq:oa}.

\begin{figure}[!t]
\centering
\includegraphics[width=1.0\columnwidth]{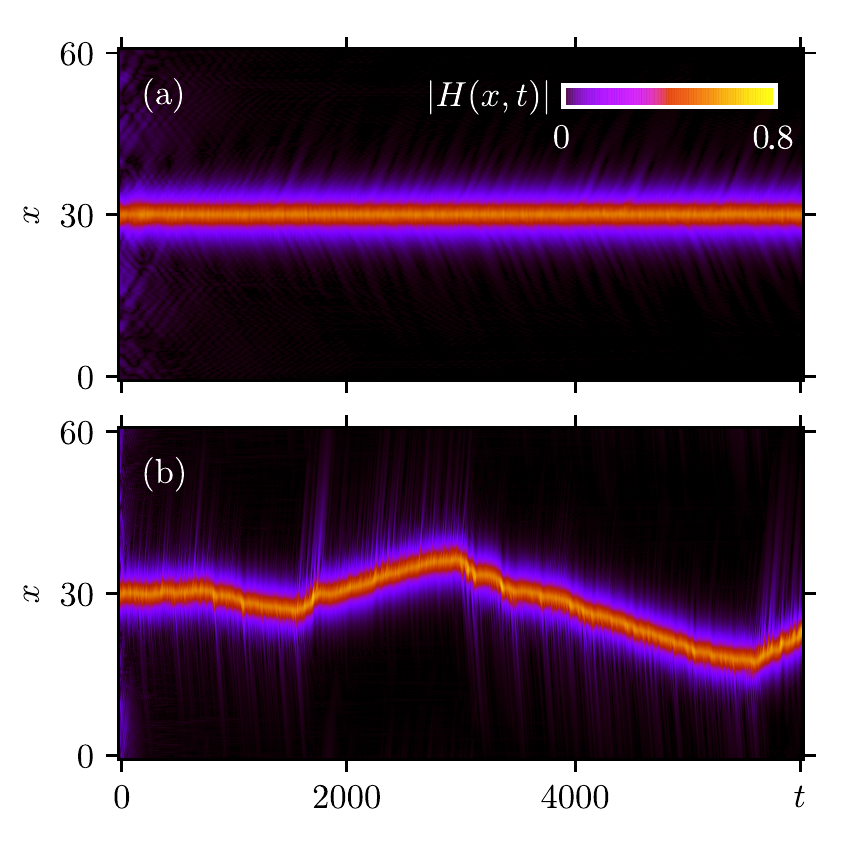}
\caption{Evolution of numerically found solitary waves in simulations
of the full equations \eqref{eq:or3} in a domain of length $L=60$. The color code
shows the field $|H(x,t)|$. Parameters: $\tau=0.5$, $\omega=0.25$, $\alpha=1.71$.
Panel (a): larger density $\rho=100$. Panel (b): small density $\rho=20$;
here one can see an onset of finite-density-induced motility at $t\approx 10^3$.}
\label{fig:solfd}
\end{figure}

\begin{figure}[t]
\centering
\includegraphics[width=\columnwidth]{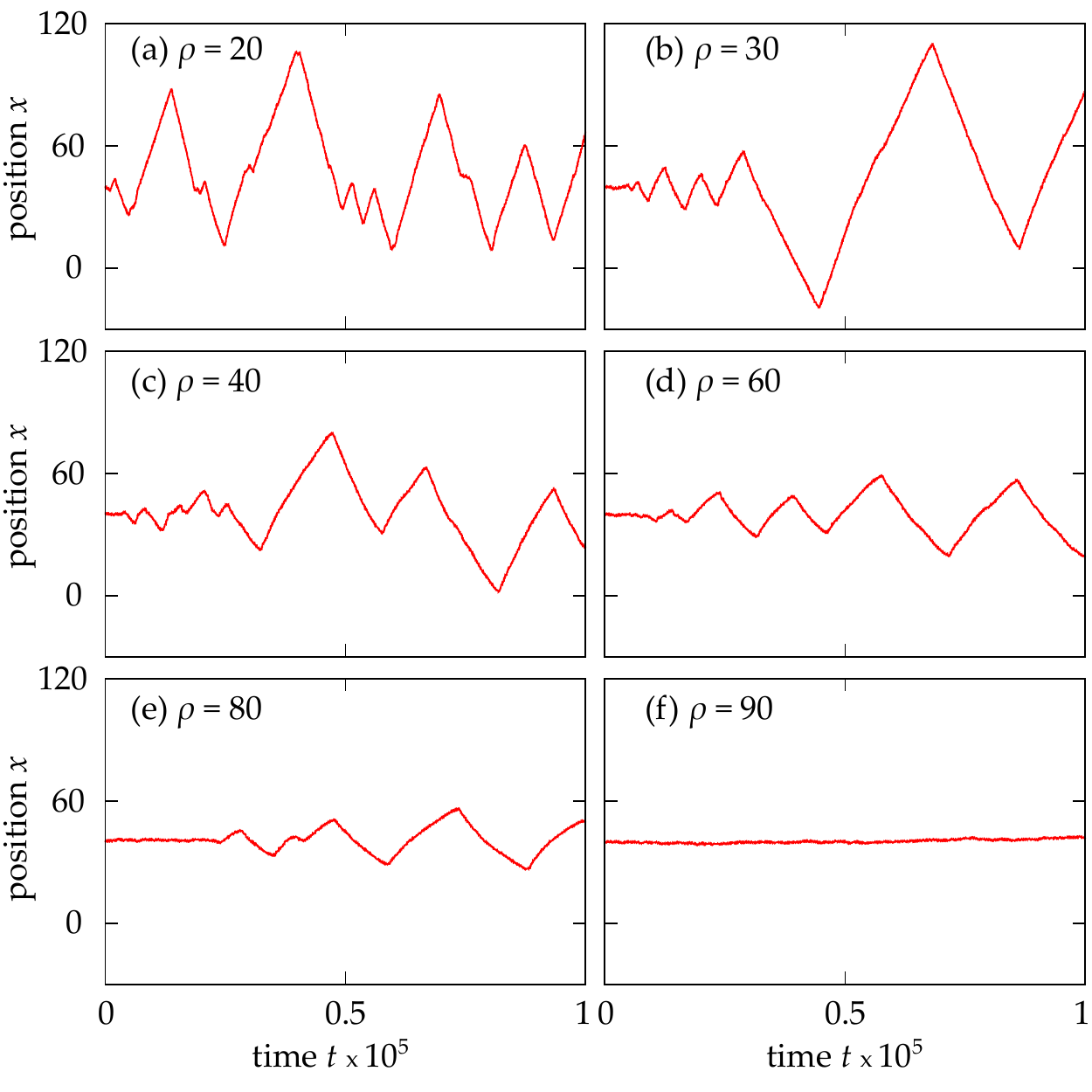}
\caption{Evolution of a soliton in a domain of size $L=80$ (this domain is ``unwrapped'' for better visibility)
for different densities $\rho$: $(\mathrm{a}){\,}\rho=20$, $(\mathrm{b}){\,}\rho=30$, $(\mathrm{c}){\,}\rho=40$,
$(\mathrm{d}){\,}\rho=60$, $(\mathrm{e}){\,}\rho=80$, and $(\mathrm{f}){\,}\rho=90$.
The line shows position of the maximum (over
$x$) of field $|H(x,t)|$. Parameters
of Eqs.~\eqref{eq:or3}: $\alpha=1.71$, $\tau=0.5$, $\omega=0.25$. While after an initial
waiting time the soliton mostly sways back and forth, several relatively short epochs during which the soliton 
stays can be seen in panels $(\mathrm{a})$, $(\mathrm{c})$, and $(\mathrm{e})$
for $\rho=20$, $\rho=40$, and $\rho=80$, respectively.}
\label{fig:mot1}
\end{figure}

We present simulations of a soliton for different densities
in Figs.~\ref{fig:solfd},{\,}\ref{fig:mot1}. The initial conditions has been chosen according to the solution
of continuous equations $Z(x)$, $H(x)$ as follows: (i) the field $H(x)$ was just taken at discrete points $x_n$;
the values $Z(x_n)$ where used to randomly sample the phases $\varphi_n$ according to
the wrapped Cauchy distribution with order parameter  $Z(x_n)$. 
Fig.~\ref{fig:solfd} show evolution on a relatively short time interval. It demonstrates that
a soliton is  robust also in simulations with a finite density.

However, for $\rho=20$ fluctuations are definitely larger than for $\rho=100$.
Such fluctuations lead to a rather surprising observation
(see Fig.~\ref{fig:solfd}{\,}(b), and Fig.~\ref{fig:mot1} for more details):
after a long initial transient epoch, during which the soliton practically
does not move (apart from small fluctuations), the soliton starts to move
with a nearly constant velocity, and from time to time changes the direction
of the motion while the speed remains nearly the same.
We call this regime \textit{finite-density-induced motility} because as we will argue below,
it disappears in the limit $\rho\to\infty$. To this end, in Fig.~\ref{fig:mot1}
we show the results of direct numerical simulations of Eqs.~\eqref{eq:or3}
on a long time interval (significantly longer than one in Fig.~\ref{fig:solfd}).
Here, the position of the maximum of absolute value $|H(x,t)|$
of the complex forcing field is depicted.
One can see for large densities, the ``waiting time'' becomes large,
and in the run with $\rho=90$, no motility is observed up to time $10^5$.

It appears that the ``swaying soliton'' has in fact three metastable states: two of motion with a nearly
constant velocity in opposite directions, and one where its velocity is very small and 
its dynamics is close to a slow diffusion.
The latter state is more probable to appear for small densities, for example for $\rho=20$ in Fig.~\ref{fig:mot1}
one can see several events where the soliton stops, but then after a relatively short
waiting time starts to sway again.

\begin{figure}[t]
\centering
\includegraphics[width=\columnwidth]{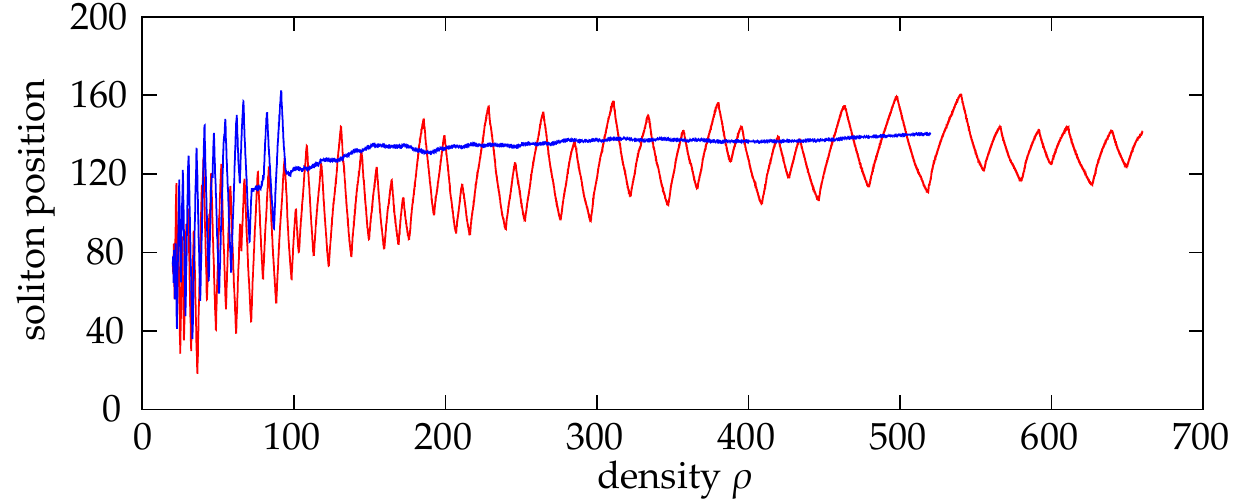}
\caption{Two runs of a soliton in a domain of size $L=150$ with continuously growing density (see description
of the procedure in the text). In one run (red) the soliton continues to move at large densities,
while in another run (blue) it at $\rho\approx 90$ spontaneously stops.}
\label{fig:add}
\end{figure}

A natural question arises, if there is a threshold in the value of the density $\rho$, 
below which the solitons can become motile, 
and above which they stay (or slowly diffuse). To check this, we performed simulations where we 
``adiabatically'' increased the density. Namely, in simulations we started with $N=3000$ oscillators
on the lattice of length $L=150$ (i.e. the initial density is $\rho=20$). Then, we added one particle 
every $50$ units of time. A particle was added in the disordered region, opposite to the synchronous domain.
In this way, the density parameter slowly grows in time while the soliton is only weakly disturbed.
Two such runs are presented in Fig.~\ref{fig:add}. In one run (blue curve), the soliton stopped 
spontaneously at density $\rho\approx 90$ and then never became motile again. But in another run (red curve)
the soliton continued to move up to densities as large as $\rho=650$ (the total simulation time 
of this run is $4.8\times 10^6$). This result suggests that there possibly is no upper threshold in density
for the soliton motility, or at least this threshold is larger than $\rho=650$.

\begin{figure}[t]
\centering
\includegraphics[width=\columnwidth]{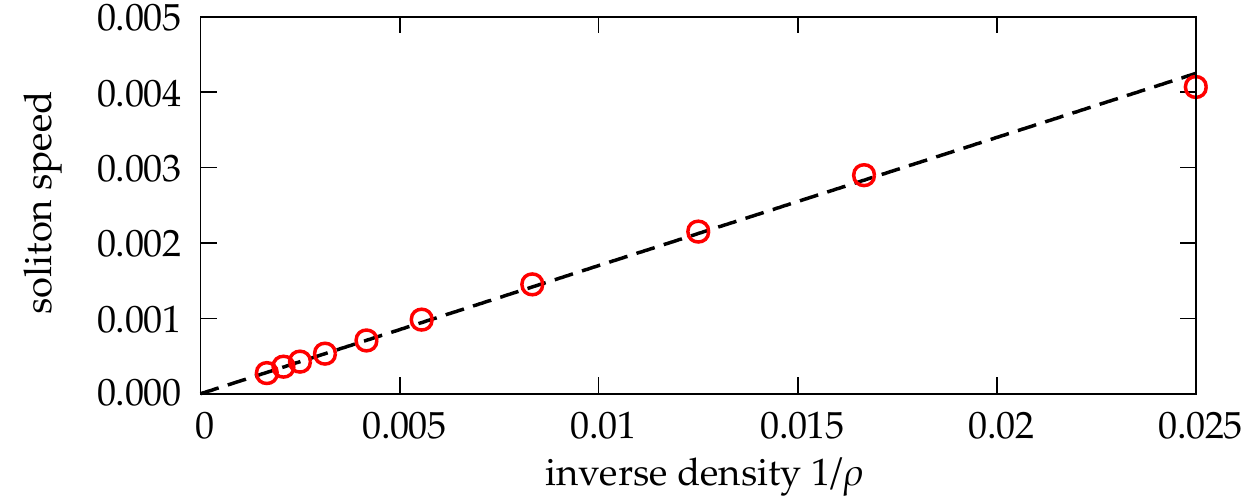}
\caption{Dependence of the soliton velocity on the density. The dashed line is the fit $V=0.17/\rho$.}
\label{fig:vr}
\end{figure}

One can already see from Fig.~\ref{fig:add} that the soliton velocity decreases with density $\rho$.
To quantify this, we have taken motile states at different stages of the red run in Fig.~\ref{fig:add},
and continued them -- but now without adding particles, i.e., at a constant density. 
Now from several stretches between U-turns
one can evaluate the average velocity of the soliton. The data shown in Fig.~\ref{fig:vr} show that with a
good accuracy this velocity is inverse proportional to the density. This means that the soliton motility 
is a pure finite-size effect, disappearing in the continuous limit $\rho\to\infty$.

\begin{figure}[t]
	\centering
	\includegraphics[width=\columnwidth]{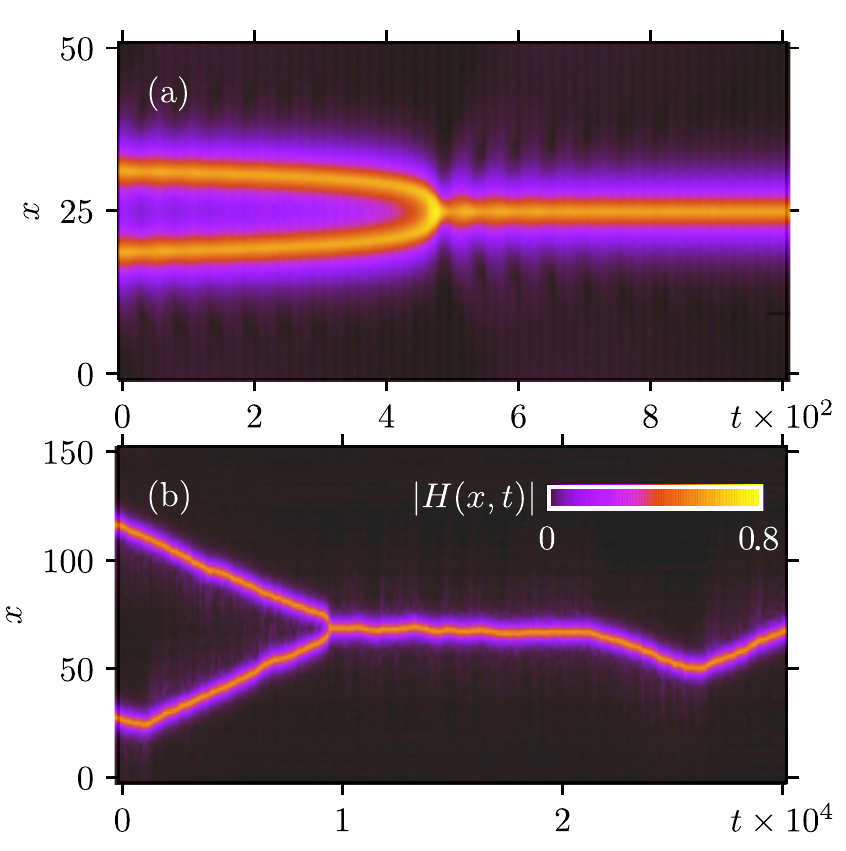}
	\caption{
		Panel (a): Space-time plot of field $|H(x,t)|$ showing merging of two solitons in  
		Eqs.~\eqref{eq:oa} with parameters
		$L=50$, $\alpha=1.71$, $\tau=0.5$, $\omega=0.25$,
		and initial distance between solitons $D=12$.	
		Panel (b): Collision and merging (at $t\approx 9500$) of two motile solitons at finite
		density $\rho=40$.
		Parameters of numerical simulation within Eqs.~\eqref{eq:or3}: $L=150$, $\alpha=1.71$, $\tau=0.5$, and $\omega=0.25$.
		At $t\approx 22000$ the soliton starts to move again. The color code shows the value of field $|H(x,t)|$.}
	\label{fig:merg+t2s}
\end{figure}

Below, we discuss possible mechanisms of the finite-size-induced motility. It appears,
that they are due to fluctuations of the field $H(x,t)$. Indeed, in the continuous limit
(i.e., in the framework of the analysis based on the OA ansatz 
in Sec.~\ref{sec:solchim}) this field is uniformly rotating, but otherwise constant. 
Synchronous oscillators are locked by this field.
Non-synchronized oscillators 
in this constant field perform rotations. These periodic motions have a zero Lyapunov
exponent (which corresponds to neutrality with respect to phase shifts). For a finite density,
the fields $H(x,t)$ starts to fluctuate. Because the fluctuations at neighboring
sites (say, $n$ and $n+1$) are highly correlated, one can in
the first approximation consider these fluctuations as common for the neighbor oscillators.
Thus one expects that periodic rotations can be synchronized by common noise (see 
Refs.~\cite{Pikovsky-84a,Pikovsky-Rosenblum-Kurths-01,Goldobin-Pikovsky-04,Goldobin-Pikovsky-05b}
for a theory of synchronization of oscillators
by common noise). For a chimera pattern with a relatively
small density of oscillators, this synchronization has been recently reported 
in Ref.~\cite{Pikovsky-21b} for a social-type setup of the network, where fluctuations are especially
enhanced and correlations of these fluctuations at neighboring oscillators are especially high.
See also Ref.~\cite{Zhang-Motter-21}, where a potential role of common-noise-type fluctuations in a formation
of chimeras has been discussed.
In the present setup, it is difficult to characterize effective fluctuations of the fields $H(x,t)$
and their correlations, due to non-stationarity effects. Nevertheless, a rough estimation
that fluctuations intensity decrease with density as $\sim \rho^{-1}$ corresponds
to the observed dependence of the soliton velocity (Fig.~\ref{fig:vr}). The effect of fluctuations
can be qualitatively described as follows: close to the synchronous domain (center of the soliton)
the fluctuations are especially strong, and they after a long action lead to a partial synchronization
of an asynchronous domain close to the soliton center. The central synchronous domain merges with this newly
created partial synchronous one and the soliton shifts its position. This picture, however, cannot explain
neither regularity and persistence of the motility, nor the U-turns. A more detailed exploration
of the soliton dynamics is needed to clarify these issues.
In Section~\ref{sec:st} (see Fig.~\ref{fig:merg+t2s}) we also demonstrate what happens when 
two moving solitons propagating in opposite directions collide.

Finally, we mention that solitons changing direction of their velocity have been previously
studied in the context of conservative nonlinear differential equations and 
lattices~\cite{Calogero-Degasperis-76,Calogero-Degasperis-05,Katz-Givli-20}. There,
solitons that just once change the direction of their motion are called boomerons, and solitons 
that periodically sway around a certain position are called trappons. Our setup differs from these studies in
several aspects. First, our soliton is a dissipative one, and we are not aware of any boomerons and trappons
in dissipative equations. Second, swaying of solitons in our case appears as a random process, not
as a regular periodic swaying.

\section{Soliton turbulence}
\label{sec:st}
In this section we describe irregular dynamics of the oscillatory medium in the situations where 
the fully incoherent state $Z=H=0$ becomes unstable. We will focus on the properties
of the system of partial differential equations \eqref{eq:oa}, i.e. we do not consider any
finite-density effects.
Correspondingly, the numerical simulations in this section are mainly concentrated on 
numerical solutions of Eqs.~\eqref{eq:oa}.
Below we will fix parameters
$\tau=0.5$ and $\omega=0.25$, exploration of other parameter values have revealed
that the described below properties are qualitatively quite general. For these parameters,
the  incoherent state $Z=H=0$ is unstable below $\alpha_l\approx 1.69$ (cf. Fig.~\ref{fig:om}), and stable
above this value.

\subsection{Merging of solitons}
\begin{figure}[t]
\centering
\includegraphics[width=\columnwidth]{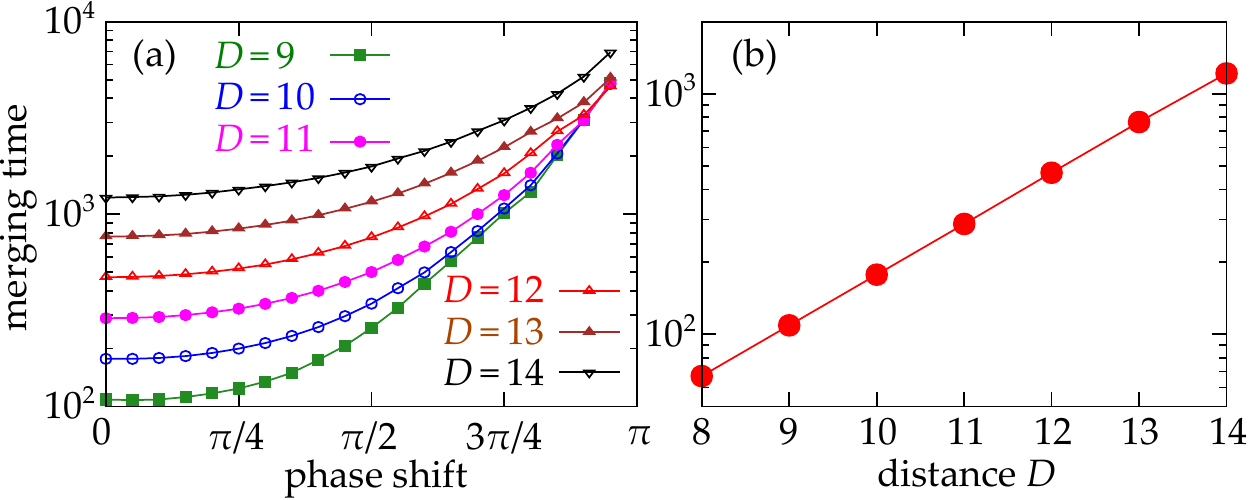}
\caption{Merging time for two solitons in dependence on the initial phase shift (a)
and on the initial distance $D$ ((b), for a zero phase shift). Parameters of simulation:
$L=60$, $\alpha=1.71$, $\tau=0.5$, $\omega=0.25$.}
\label{fig:2s}
\end{figure}

We start with the value of the phase shift $\alpha=1.71$, at which a stable soliton exists.
We illustrate in Fig.~\ref{fig:merg+t2s}(a), that two solitons, placed closed to each other, 
experience attraction and eventually merge.
The interaction is, however, dependent on the initial phase shift between the solitons.
As illustrated in Fig.~\ref{fig:2s}(a), the merging time is minimal if the phase shift is zero, 
and grows if the phase shift becomes close to $\pi$. Because the soliton has exponentially decaying tails,
one can expect that the interaction decreases exponentially with the initial separation
distance $D$. This is confirmed in Fig.~\ref{fig:2s}(b), where we show the exponential 
dependence of the merging time on the initial distance.

 In the simulation, performed within the system~\eqref{eq:or3} and
depicted in Fig.~\ref{fig:merg+t2s}(b), we also explore what happens when two
motile solitons, moving in opposite directions, collide.
One can see that they merge and produce one stationary soliton,
similar to the case of soliton merging in the infinite-density limit described above.
This soliton, however, after a certain waiting time, starts to move.

\subsection{Creation and merging of solitons close to instability border}
At $\alpha=\alpha_l\approx 1.69$, the incoherent state $Z=H=0$ becomes unstable,
this instability is a long-wave one (cf. Fig.~\ref{fig:om}). Below in Fig.~\ref{fig:ff1} we explore the result 
of this instability. 
Panels (a,b,c) show the regimes close to the threshold $\alpha_l$.
One observes events of merging of solitons due to their attractive interaction,
like in Fig.~\ref{fig:merg+t2s}(a). In created voids, new solitons appear due to the long-wave instability.
The whole irregular process in Fig.~\ref{fig:ff1}(a,b,c) can be characterize as a
repeating creation and merging of solitons.
The characteristic spatial scale diverges close to the criticality $\alpha_l$, and correspondingly diverges
the characteristic time scale (according to the exponential law of Fig.~\ref{fig:2s}). This slowing down 
of the process close to criticality is
clearly seen if one compares time scales on panels (a-c). 

\begin{figure*}[!htb]
\centering
\includegraphics[width=\textwidth]{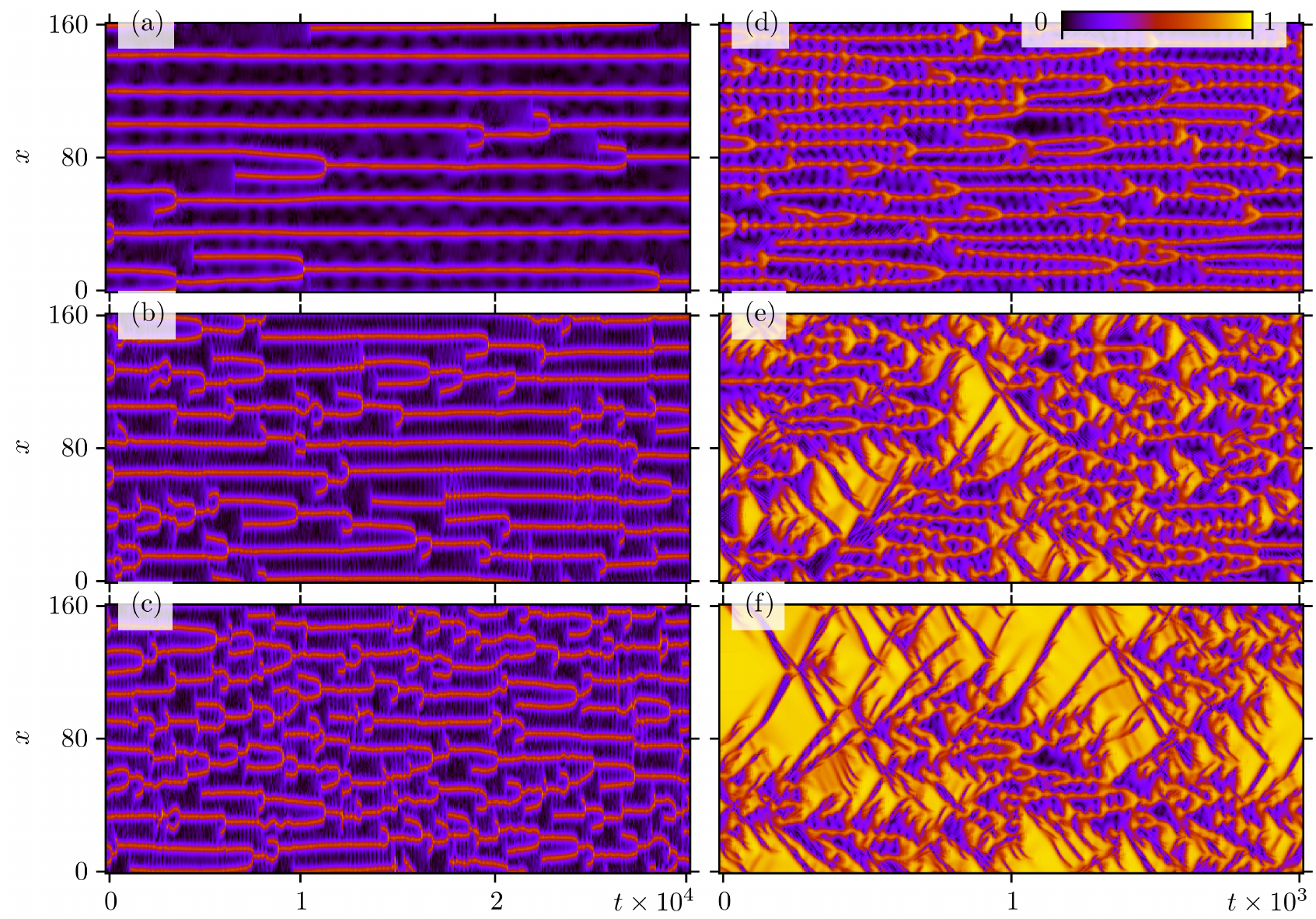}
\caption{Spatio-temporal plots of field $|H(x,t)|$ for  different
regimes of soliton turbulence for $\tau=0.5$, $\omega=0.25$.
Panel (a): $\alpha=1.69$; panel (b):  $\alpha=1.685$; panel (c):  $\alpha=1.68$;
panel (d): $\alpha=1.65$; panel (e):  $\alpha=1.6$; panel (f):  $\alpha=1.59$.}
\label{fig:ff1}
\end{figure*}

\subsection{Transition to spatio-temporal intermittency}
At $\alpha=1.65$ 
(relatively far away from the threshold) one still can characterize the process as creations and 
annihilations of solitons (Fig.~\ref{fig:ff1}(d)).
Here, the characteristic space and time scale further decrease (the time scale in panel (d)
is reduced compared to panel (c) by factor ten). 
However, in panel (d) one observes an additional feature at the merging events.
Because at the merging a wide region of synchrony appears, and for smaller $\alpha$ the 
stability of the synchronous state $|Z|=1$ is enhanced, relatively large and long-living patches 
of synchrony appear, they can be recognized as bright triangle-shaped regions in Fig.~\ref{fig:ff1}(d).
Nevertheless, the surrounding incoherence eventually wins so that these synchronous
patches remain small and isolated. With further decrease of parameter $\alpha$,
synchronous patches become larger and long-living (panels (e,f) in Fig.~\ref{fig:ff1}). 
This regime can be characterized as a spatio-temporal intermittency, and
this intermittency becomes more pronounced for smaller values of $\alpha$.
Here the synchronous patches (bright yellow regions in panel (f))
dominate, but still coexist with incoherent domains. In many cases one observes
narrow incoherent pulses propagating on a synchronous background. A detailed description of
such ``dark solitons'' will be presented elsewhere.

It should be stressed that for the values of $\alpha<\alpha_l$, the fully synchronous state
$|Z|=1$ is stable, and thus it is an absorbing state for the spatio-temporal intermittency. 
We have observed that below the critical value $\alpha_{sti}\approx 1.58$, the absorbing state
wins and the spatio-temporal intermittency is a transient. Here the final stable regime 
in the oscillatory medium is full synchrony.

\section{Conclusions}
\label{sec:concl}
Here we highlight essential novel findings of this paper. 

First, we extended the classical Kuramoto-Battogtokh model of coupled oscillators to the case of a finite inertia
of the diffusive field responsible for the coupling. This extension allows for a consideration of more realistic
physical situations compared to the original KB formulation, which required a strong time scale
separation between the oscillators period and the characteristic relaxation time of the diffusive field. Remarkably,
together with the time scale of diffusion, the frequency of the oscillators now is a relevant parameter, which
cannot be ``killed'' by a transformation to a rotating reference frame. At this point we would like to mention, that
there have been numerious studies of chimera patterns in non-locally coupled oscillators, but with kernels
of coupling different from the exponential kernel adopted by Kuramoto and Battogtokh. Such kernels
(for example, a periodic in space $\cos$-shaped kernel or a rectangular kernel) may have some
advantages in mathematical and/or computational treatment of the equations, but they are not as 
physically motivated as the exponential kernel originated from a diffusion process.
Correspondingly, to the best of our knowledge, there are no realistic extensions of such kernels to the case of a 
finite time scale of the propagation of the coupling.

In the model with a finite time scale of diffusion, we have found stable chimera solitons. 
They exist for phase shifts in coupling, where there is a bistability of fully synchronous and fully asynchronous 
spatially homogeneous states. Because the tails of solitons decay exponentially, they are non-sensitive to boundary
conditions if the medium length is significantly larger than the soliton width.  Thus, for solitons, 
in contradistinction
to patterns in the KB model, we do not have to impose
periodic boundary conditions. 

We have found soliton solutions numerically in partial differential 
equations describing the system in a continuous limit of an infinitely high density of oscillators.
These equations appear in the standard Ott-Antonsen approach, where a coarse-grained order parameter is 
introduced. Soliton solutions have been found numerically as homoclinic trajectories 
of the resulting three-dimensional 
system of ordinary differential equations for the spatial profile. 
In addition, we have developed an approximate analytical
theory, using the fact that at certain parameters a soliton having either empty or a single point synchronous domain  
can be found analytically as a solution of a second-order equation. We have demonstrated that
this analytical first-order perturbation theory provides a good approximation in 
a certain range of parameters.

The most surprizing finding is the finite-size-induced motility of solitons. 
It is always present in
microscopic  simulations  with
a finite (and relatively small) density of oscillators. We demonstrated that this motility leads to an irregular swaying of
a soliton. Typically, the dynamics has three states with velocities $\approx\pm V$ and $\approx 0$, while
the latter regime in most cases appears seldom and for a small time interval. In our discussion we attributed
the motility to finite-size fluctuations of the diffusive field at the central part of the soliton.
Our hypothesis is that these fluctuations partially synchronize neighboring states, due to the effect
of synchronization by common noise, and thus the soliton moves. This picture is consistent with the observed
dependence of the characteristic velocity on the density $V\sim \rho^{-1}$. Nevertheless,
there are many open questions that should be addressed in future studies, e.g. why motion of solitons
is so regular and why they from time to time reverse the direction of motion. At this point we would like to mention,
that there are not so many dissipative nonlinear systems allowing for both microscopic and macroscopic descriptions.
Apart from coupled oscillators, one can mention granular systems~\cite{Aranson-Tsimring-09}. There, however,
the interactions are local and thus one can hardly expect finite-size effects of synchronization 
through fluctuations, like
in the model considered in this paper.

We demonstrated that chimera solitons interact attractively, so two solitons placed at some distance
come close to each other and eventually merge. The interaction is less attractive if the solitons
are prepared with a phase shift $\approx \pi$, in this case during a long initial stage they first 
adjust their phases and then merge.

Merging of solitons is the essential process shaping the soliton turbulence, which is
observed when the asynchronous  homogeneous state in the medium becomes unstable.
This long-wave instability leads to appearance of a ``chain of solitons'', which, however is not stable.
On a long time scale, mergings occur. These merging leave larger domains of asynchrony where, 
do to instability, new
solitons emerge. Thus, the whole dynamics can be described as a sequence of irregular 
events of merging and emerging of solitons. Because the solitons are the basic constituents 
of the observed pattern, it is natural to speak on soliton turbulence in this context. 
Furthermore, we have followed the development of the soliton turbulence in the domain 
of parameters, where the stability of the synchronous state is stronger. Here, the 
extended synchronous patches that appear at the merging of solitons, become long-living. 
With further increase of stability of synchronous states the pattern ``reverses'': 
instead of localized synchronous objects on the asynchronous background, one observes 
large synchronous domains with relatively small asynchronous  patched between them. 
This regime can be characterized as spatio-temporal intermittency, where the 
synchronous state is absorbing.

\begin{acknowledgments}
This paper was supported by the Russian 
Science Foundation (Sec.~\ref{sec:bm}, \ref{sec:hss}, \ref{sec:solchim}, Grant No.~19-12-00367) and the
Ministry of Science and Higher Education of the Russian
Federation (Sec.~\ref{sec:fsim}, \ref{sec:st}, Grant No.~0729-2021-013 (BK-P/23, date 14.09.2021)).
A.P. acknowledges support by DFG (Grant PI 220/22-1).
We thank M. Wolfrum and O. Omelchenko for fruitful discussions.
\end{acknowledgments}

%

\end{document}